\documentclass[11pt]{article}
\usepackage{geometry}   
\usepackage{dsfont}
\usepackage{amsthm}
\usepackage{amsmath}
\usepackage{amssymb}
\usepackage{esint}
\usepackage{graphicx}
\usepackage{mathrsfs}             
\usepackage{graphicx}
\usepackage{amssymb}
\usepackage{epstopdf}
\usepackage{textpos} 
\title{Bell Tests for Histories}
\author{Jordan Cotler$^1$ and Frank Wilczek$^{1,2}$\\
\small\it 1. Center for Theoretical Physics, MIT, Cambridge MA 02139 USA \\
\small \it 2. Origins Project, Arizona State University, Tempe AZ 25287 USA}

\begin{document}
\maketitle

\begin{textblock*}{5cm}(11cm,-8.2cm)
  \fbox{\footnotesize MIT-CTP-4653}
\end{textblock*}

\begin{abstract}
We describe a procedure to create entangled history states and measurements that would enable one to check for temporal entanglement.  The checks take the form of inequalities among observable quantities.  They are similar in spirit, but different in detail, to Bell tests for ordinary entanglement.
\end{abstract}

In a previous paper \cite{entangledHistories} we constructed a framework for considering history states and history observables of quantum mechanical systems.   A novel possibility arising in that framework is the existence of entangled histories.  Here we demonstrate that the concept of entangled histories involves testable consequences, in the form of Bell-like inequalities.  We will carry over definitions and notations from \cite{entangledHistories}.



\section{Constructing an Entangled History State}

To begin, consider a spin-1/2 particle in the state $|x^+\rangle = \frac{1}{\sqrt{2}} (|z^+\rangle + |z^-\rangle)$ and two auxiliary qubits.     
We introduce two auxiliary qubits $|0\rangle_1 |0\rangle_2 \equiv |00\rangle$.   At time $t_1$ we perform a CNOT operation between the first auxiliary qubit and the spin-1/2 particle, resulting in
\begin{equation}
\frac{1}{\sqrt{2}} \,|z^+\rangle|00\rangle + \frac{1}{\sqrt{2}} \,|z^-\rangle |10\rangle
\end{equation}
We let this system evolve trivially to time $t_2$.  Then at time $t_2$, we perform a CNOT between the second auxiliary qubit and the spin-1/2 particle, resulting in
\begin{equation}
\frac{1}{\sqrt{2}} \,|z^+\rangle|00\rangle + \frac{1}{\sqrt{2}} \,|z^-\rangle |11\rangle
\end{equation}

If we measure the auxiliary qubits in the $\{|00\rangle, |11\rangle,...\}$ basis, then measuring $|00\rangle$ would indicate that the spin-1/2 particle has been in the history state $[z^+] \odot [z^+]$; and if we measure $|11\rangle$, this would indicate that the spin-1/2 particle has been in the history state $[z^-] \odot [z^-]$.  However we can also choose to measure the auxiliary qubits in the Bell basis $\left\{\frac{1}{\sqrt{2}}(|00\rangle \pm |11\rangle),...\right\}$.  Then if we measure $\frac{1}{\sqrt{2}}(|00\rangle + |11\rangle)$, it means that the spin-1/2 particle has been in the history state $[z^+] \odot [z^+]$ with amplitude $1/\sqrt{2}$, \textit{and} $[z^-] \odot [z^-]$ with amplitude $1/\sqrt{2}$.  In other words, the particle has been in the entangled history state $\frac{1}{\sqrt{2}} \left( [z^+] \odot [z^+] + [z^-] \odot [z^-] \right)$.  By changing the basis of the auxiliary qubits, we have \textit{erased} knowledge about the history of the spin-1/2 particle.  As emphasized in \cite{e2i2}, selective erasure can be a powerful tool for exploring quantum interference phenomena.


\section{Temporal CHSH Inequality}

\subsection{Standard CHSH Inequality}
Before considering the temporal CHSH inequality, let us first review the standard CHSH inequalities.  We will focus our attention on two spin-1/2 particles at a single fixed time.  Let the quantum state of the particles be $|\Psi\rangle$.  Letting
\begin{equation}
|\chi(\theta, \phi)\rangle = \begin{bmatrix}
\cos \theta \\
e^{i \phi} \, \sin \theta
\end{bmatrix}, \qquad |\chi_\perp(\theta, \phi)\rangle = \begin{bmatrix}
- e^{-i \phi}\,\sin \theta \\
\cos \theta
\end{bmatrix}
\end{equation}
in the $\{|z^+\rangle, |z^-\rangle\}$ basis, we define
\begin{align}
E_\Psi(\theta_1, \phi_1 ; \theta_2, \phi_2) \, \equiv \,\,& \left|\bigg(\langle \chi(\theta_1, \phi_1) |\otimes \langle \chi(\theta_2, \phi_2)|\bigg)|\Psi\rangle \right|^2 -\left|\bigg(\langle \chi_\perp(\theta_1, \phi_1) |\otimes \langle \chi(\theta_2, \phi_2)|\bigg)|\Psi\rangle \right|^2 \nonumber \\
&-  \left|\bigg(\langle \chi(\theta_1, \phi_1) |\otimes \langle \chi_\perp(\theta_2, \phi_2)|\bigg)|\Psi\rangle \right|^2 +\left|\bigg(\langle \chi_\perp(\theta_1, \phi_1) |\otimes \langle \chi_\perp(\theta_2, \phi_2)|\bigg)|\Psi\rangle \right|^2
\end{align}
and also
\begin{equation}
S_\Psi(\theta_1, \phi_1; \theta_2, \phi_2; \theta_3, \phi_3; \theta_4, \phi_4) \,  =  \, E_\Psi(\theta_1, \phi_1; \theta_2, \phi_2) - E_\Psi(\theta_1, \phi_1; \theta_4, \phi_4) + E_\Psi(\theta_3, \phi_3; \theta_2, \phi_2) + E_\Psi(\theta_3, \phi_3; \theta_4, \phi_4)
\end{equation}
The CHSH inequality \cite{CHSH} tells us that if our quantum particle behaves like a classical particle (possibly with hidden variables), then
\begin{equation}
-2 \leq S \leq 2
\end{equation}
However, for the entangled Bell state $|\Psi\rangle = \frac{1}{\sqrt{2}} (|z^+ z^+\rangle + |z^- z^-\rangle)$, we have
\begin{equation}
S_{\Psi}(0,0;\pi/8,0;\pi/4,0; 3\pi/8,0) = 2 \sqrt{2}
\end{equation}
which violates the CHSH inequality.  Tsirelson's bound \cite{tsirelson} tells us that $2 \sqrt{2}$ is a maximal violation of the CHSH inequalities.
\subsection{Temporal Variation}
There is a natural generalization of the CHSH inequality to a single spin-1/2 particle at two times $t_1 < t_2$ with bridging operator $T(t_2, t_1)$.  Suppose that the initial state of the particle is $|\Psi(t_1)\rangle$.  We project this state onto $|\chi(\theta_1, \phi_1)\rangle$, and let the residual state evolve to $T(t_2, t_1)|\chi(\theta_1, \phi_1)\rangle$ at time $t_2$.  Then at time $t_2$, we project onto $|\chi(\theta_2, \phi_2)\rangle$.  Thus, at the end of the procedure, we are left at time $t_2$ in the residual state $|\chi(\theta_2, \phi_2)\rangle$ with probability
\begin{equation}
\bigg|\langle \chi(\theta_2, \phi_2)|T(t_2, t_1)|\chi(\theta_1, \phi_1)\rangle \langle \chi(\theta_1, \phi_1)| \Psi(t_1)\rangle \bigg|^2
\end{equation}

Let us define
\begin{align}
\label{Etilde}
\widetilde{E}_\Psi(\theta_1, \phi_1 ; \theta_2, \phi_2) := \,\,& \bigg|\langle \chi(\theta_2, \phi_2)|T(t_2, t_1)|\chi(\theta_1, \phi_1)\rangle \langle \chi(\theta_1, \phi_1)| \Psi(t_1)\rangle \bigg|^2 \nonumber \\
&- \bigg|\langle \chi(\theta_2, \phi_2)|T(t_2, t_1)|\chi_\perp(\theta_1, \phi_1)\rangle \langle \chi_\perp(\theta_1, \phi_1)| \Psi(t_1)\rangle \bigg|^2 \nonumber \\
&-  \bigg|\langle \chi_\perp(\theta_2, \phi_2)|T(t_2, t_1)|\chi(\theta_1, \phi_1)\rangle \langle \chi(\theta_1, \phi_1)| \Psi(t_1)\rangle \bigg|^2 \nonumber \\
&+ \bigg|\langle \chi_\perp(\theta_2, \phi_2)|T(t_2, t_1)|\chi_\perp(\theta_1, \phi_1)\rangle \langle \chi_\perp(\theta_1, \phi_1)| \Psi(t_1)\rangle \bigg|^2
\end{align}
and also
\begin{equation}
\label{sTilde}
\widetilde{S}_\Psi(\theta_1, \phi_1; \theta_2, \phi_2; \theta_3, \phi_3; \theta_4, \phi_4) := \widetilde{E}_\Psi(\theta_1, \phi_1; \theta_2, \phi_2) - \widetilde{E}_\Psi(\theta_1, \phi_1; \theta_4, \phi_4) + \widetilde{E}_\Psi(\theta_3, \phi_3; \theta_2, \phi_2) + \widetilde{E}_\Psi(\theta_3, \phi_3; \theta_4, \phi_4)
\end{equation}
Time evolution of a classical system implies 
\begin{equation}
\label{tildeInequality}
-2 \leq \widetilde{S} \leq 2
\end{equation}
which is known as the temporal CHSH inequality \cite{temporalCHSH, fritz1, LGreview}.  

Interestingly, very simple quantum systems violate this temporal CHSH ($\tau$CHSH) inequality.  For example, a spin-1/2 particle beginning in the state $|\Psi(t_1)\rangle = |z^+\rangle$ with trivial bridging operator $T = \textbf{1}$ satisfies
\begin{equation}
\widetilde{S}_{\Psi}(0,0;\pi/8,0;\pi/4,0; 3\pi/8,0) = 2 \sqrt{2}
\end{equation}
which saturates a temporal analog of the Tsirelson bound.  Here the $\tau$CHSH inequality is detecting deviations from unitary evolution due to the projection operator representing measurements.   In this sense, quantum measurement is a non-classical process.

Here we are less interested in this (effectively) non-unitary evolution than in temporal entanglement.  But since systems with trivial history structure can maximally violate the $\tau$CHSH inequality, we need a different criterion to distinguish entangled histories.  



Before identifying such a criterion, we will first consider a more sophisticated treatment of history states that does a better job of compensating for the projection that accompanies measurement.

\subsection{Temporal CHSH Inequality for $[z^+]\odot[z^+]$}

First, let us show how to apply the temporal CHSH inequality to the non-entangled history state $[z^+]\odot[z^+]$.  We will show that in our more sophisticated treatment this history state does \textit{not} lead to violation of the $\tau$CHSH, Eqn.\,(\ref{tildeInequality}).  To obtain a version of the various terms appearing in Eqn. (\ref{Etilde}) that is faithful to that history, we would like to be able to project the history state $[z^+]\odot[z^+]$ onto some $|\chi(\theta, \phi)\rangle$ or $|\chi_\perp(\theta, \phi)\rangle$ at time $t_1$, and then project onto some $|\chi(\theta', \phi')\rangle$ or $|\chi_\perp(\theta', \phi')\rangle$ at time $t_2$.

For concreteness, say that we want to project onto $|\chi(\theta, \phi)\rangle$ at time $t_1$ and then onto $|\chi_\perp(\theta', \phi')\rangle$ at time $t_2$.  The simplest strategy is to prepare our spin-1/2 particle in the initial state $|z^+\rangle$ at time $t_1$, and then apply $|\chi(\theta, \phi)\rangle \langle \chi(\theta, \phi)|$ which leaves us in the state
\begin{equation}
\bigg( \langle \chi(\theta, \phi)|z^+\rangle \bigg) |\chi(\theta, \phi)\rangle
\end{equation}
This state evolves trivially to time $t_2$.  However, the resulting state at time $t_2$ is \textit{not} $|z^+\rangle$ as is required by the history state $[z^+]\odot[z^+]$.  To remedy that issue, we rotate $| \chi(\theta, \phi)\rangle \to |z^+\rangle$ at time $t_2$, leaving us in the state
\begin{equation}
\bigg( \langle \chi(\theta, \phi)|z^+\rangle \bigg) |z^+\rangle
\end{equation}
Then we can project onto $|\chi(\theta',\phi')\rangle$, leaving us in the state $|\chi(\theta',\phi')\rangle$ with a total probability
\begin{equation}
\label{finalprob1}
|\langle \chi(\theta', \phi')|z^+\rangle \langle \chi(\theta, \phi)|z^+\rangle|^2
\end{equation}
We were able to obtain Eq. (\ref{finalprob1}) for the history state $[z^+]\odot[z^+]$ by applying a rotation to undo the influence of the first projection, a procedure that can be realized in experiments.

From terms like the one in Eq. (\ref{finalprob1}), we can build up $\widetilde{E}_{[z^+]\odot [z^+]}(\theta_1, \phi_1; \theta_2, \phi_2; \theta_3, \phi_3; \theta_4, \phi_4)$ and $\widetilde{S}_{[z^+]\odot [z^+]}(\theta_1, \phi_1; \theta_2, \phi_2; \theta_3, \phi_3; \theta_4, \phi_4)$ for the history state $[z^+]\odot [z^+]$.  We find that
\begin{equation}
\widetilde{S}_{[z^+]\odot [z^+]}(\theta_1, \phi_1; \theta_2, \phi_2; \theta_3, \phi_3; \theta_4, \phi_4) = \cos(2\theta_1) \left(\cos(2 \theta_2) - \cos(2 \theta_4) \right) + \cos(2\theta_3) \left(\cos(2 \theta_2) + \cos(2 \theta_4) \right)
\end{equation}
and as a consequence
\begin{equation}
-2 \leq \widetilde{S}_{[z^+]\odot [z^+]} \leq 2
\end{equation}
Thus $[z^+] \odot [z^+]$ does not violate the $\tau$CHSH inequality.

\subsection{Temporal CHSH Inequality for $\frac{1}{\sqrt{2}}([z^+]\odot[z^+] + [z^-]\odot[z^-])$}

Now we will turn to the more complicated task of calculating the quantities in the $\tau$CHSH inequality for the history state $\frac{1}{\sqrt{2}}([z^+]\odot[z^+] + [z^-]\odot[z^-])$.  We will use the same strategy from the previous section.  

Again for concreteness, say that we want to project onto $|\chi(\theta, \phi)\rangle$ at time $t_1$ and then onto $|\chi_\perp(\theta', \phi')\rangle$ at time $t_2$.  We let our spin-1/2 particle start in the initial state $|x^+\rangle = \frac{1}{\sqrt{2}}(|z^+\rangle + |z^-\rangle)$ at time $t_1$, and put in two auxiliary qubits $|0\rangle_1 |0\rangle_2 = |00\rangle$.  We then perform a CNOT between the first auxiliary qubit and the spin-1/2 particle, resulting in the state
\begin{equation}
\frac{1}{\sqrt{2}} \, |z^+\rangle |00\rangle + \frac{1}{\sqrt{2}} \, |z^-\rangle |10\rangle
\end{equation}
We then project the spin-1/2 particle onto $|\chi(\theta, \phi)\rangle$ which leaves us in the state
\begin{equation}
\frac{1}{\sqrt{2}} \, \bigg(\langle \chi(\theta, \phi)|z^+\rangle \bigg) |\chi(\theta, \phi)\rangle |00\rangle + \frac{1}{\sqrt{2}} \,  \bigg(\langle \chi(\theta, \phi)|z^-\rangle \bigg) |\chi(\theta, \phi)\rangle |10\rangle
\end{equation}
which in turn evolves trivially to time $t_2$.  We then apply a controlled rotation between the first auxiliary qubit and the spin-1/2 particle, sending $|\chi(\theta, \phi)\rangle |00\rangle \to |z^+\rangle |00\rangle$ and $|\chi(\theta, \phi)\rangle |10\rangle \to |z^-\rangle |10\rangle$ resulting in the state
\begin{align}
\frac{1}{\sqrt{2}} \, \bigg(\langle \chi(\theta, \phi)|z^+\rangle \bigg) |z^+\rangle |00\rangle + \frac{1}{\sqrt{2}} \,  \bigg(\langle \chi(\theta, \phi)|z^-\rangle \bigg) |z^-\rangle |10\rangle
\end{align}
Next, we perform a CNOT operation between the second auxiliary qubit and the spin-1/2 particle yielding
\begin{align}
\frac{1}{\sqrt{2}} \, \bigg(\langle \chi(\theta, \phi)|z^+\rangle \bigg) |z^+\rangle |00\rangle + \frac{1}{\sqrt{2}} \,  \bigg(\langle \chi(\theta, \phi)|z^-\rangle \bigg) |z^-\rangle |11 \rangle
\end{align}
Projecting onto the entangled auxiliary state $\frac{1}{\sqrt{2}} (|00\rangle + |11\rangle)$ and post-selecting on this, we are left with the suitably renormalized state
\begin{equation}
\frac{1}{\sqrt{2}} \, \bigg(\langle \chi(\theta, \phi)|z^+\rangle \bigg) |z^+\rangle + \frac{1}{\sqrt{2}} \,  \bigg(\langle \chi(\theta, \phi)|z^-\rangle \bigg) |z^-\rangle
\end{equation}
where we have traced out the auxiliary qubits.  Finally, projecting onto $|\chi(\theta',\phi')\rangle$, the final state will be  $|\chi(\theta',\phi')\rangle$ with probability
\begin{equation}
\label{finalProb2}
\left|\frac{1}{\sqrt{2}}\,\langle \chi(\theta,\phi)|z^+\rangle \langle \chi(\theta',\phi')|z^+\rangle + \frac{1}{\sqrt{2}}\,\langle \chi(\theta,\phi)|z^-\rangle \langle \chi(\theta',\phi')|z^-\rangle  \right|^2
\end{equation}
taking into account the renormalization due to post-selection.  

From Equation (\ref{finalProb2}) we can construct $\widetilde{E}_{\frac{1}{\sqrt{2}}([z^+]\odot[z^+] + [z^-]\odot[z^-])}$ and $\widetilde{S}_{\frac{1}{\sqrt{2}}([z^+]\odot[z^+] + [z^-]\odot[z^-])}$.  We have
\begin{equation}
\widetilde{S}_{\frac{1}{\sqrt{2}}([z^+]\odot[z^+] + [z^-]\odot[z^-])}(0,0;\pi/8,0;\pi/4,0; 3\pi/8,0) = 2 \sqrt{2}
\end{equation}
Thus the entangled history state $\frac{1}{\sqrt{2}}([z^+]\odot[z^+] + [z^-]\odot[z^-])$ violates the $\tau$CHSH inequality.

By comparing the $\tau$CHSH inequality with the standard (non-temporal) CHSH inequality, we see that
\begin{equation}
\widetilde{S}_{\frac{1}{\sqrt{2}}([z^+]\odot[z^+] + [z^-]\odot[z^-])}(\theta_1, \phi_1; \theta_2, \phi_2; \theta_3, \phi_3; \theta_4, \phi_4) = S_{\frac{1}{\sqrt{2}}(|z^+ z^+\rangle + |z^- z^-\rangle)}(\theta_1, \phi_1; \theta_2, \phi_2; \theta_3, \phi_3; \theta_4, \phi_4)
\end{equation}
This reflects the logic of our construction, whereby the \textit{one}-particle history state $\frac{1}{\sqrt{2}}([z^+]\odot[z^+] + [z^-]\odot[z^-])$ provides a direct temporal analog of the \textit{two}-particle Bell state $\frac{1}{\sqrt{2}}(|z^+ z^+\rangle + |z^- z^-\rangle)$.

\subsection{Discussion}
While the $\tau$CHSH inequality allows for the exploration of how measured systems effectively violate unitarity, it is not a perfect metric for characterizing temporal entanglement in history states.  As we have seen, applying the $\tau$CHSH inequality to a trivially evolving particle leads to a violation of the inequality, whereas application to a trivial history state (as achieved by undoing the influence of projection) does \textit{not} lead to a violation.  On the other hand, application of the $\tau$CHSH inequality to an entangled history state (as achieved by a more elaborate post-selection scheme) \textit{does} lead to a violation of the inequality.

We desire stronger inequalities that will distinguish among the three situations we have considered in the three preceding subsections: unitary evolution of a trivial initial state, a non-entangled history state, and an entangled history state.  That is the burden of our next section.

\section{Stronger Inequalities}

For simplicity, we here assume that all time evolution is trivial.  Let $\Psi$ either denote a spin-1/2 particle evolving from time $t_1$ to $t_2$ with a fixed initial state, \textit{or} a history state of a spin-1/2 particle for the times $t_1$ and $t_2$.  In either case, we can define the operator
\begin{equation}
\label{op1}
\text{Proj}_{|\chi(\theta_1, \phi_1)\rangle, |\chi(\theta_2, \phi_2)\rangle} (\Psi)
\end{equation}
which is the probability amplitude corresponding to the projection of $\Psi$ onto $|\chi(\theta_1, \phi_1)\rangle$ at time $t_1$, and onto $|\chi(\theta_2, \phi_2)\rangle$ at time $t_2$.  Using Equation (\ref{op1}), we can define the functional 
\begin{equation}
M(\Psi) := \frac{1}{(2 \pi)^4}\int_{[0, 2\pi]^4} \left|\text{Proj}_{|\chi(\theta_1, \phi_1)\rangle, |\chi(\theta_2, \phi_2)\rangle} (\Psi) \right|^2 \, d\theta_1 \, d\phi_1 \, d\theta_2 \, d\phi_2
\end{equation}
which is the mean value of $\left|\text{Proj}_{|\chi(\theta_1, \phi_1)\rangle, |\chi(\theta_2, \phi_2)\rangle} (\Psi) \right|^2$.

For an arbitrary initial state $|\Psi(\theta, \phi)\rangle = \begin{bmatrix}
\cos \theta \\ e^{i \phi} \, \sin \theta \end{bmatrix}$ that evolves trivially in time we have
\begin{equation}
M(\text{initial state }|\Psi(\theta, \phi)\rangle) = \frac{1}{4}
\end{equation}
For an arbitrary normalized non-entangled history state
\begin{equation}
\frac{1}{|\langle \Psi(\theta', \phi')|\Psi(\theta, \phi)\rangle|}\,[\Psi(\theta', \phi')] \odot [\Psi(\theta, \phi)]
\end{equation}
we also have
\begin{equation}
M\left(\frac{1}{|\langle \Psi(\theta', \phi')|\Psi(\theta, \phi)\rangle|}\,[\Psi(\theta', \phi')] \odot [\Psi(\theta, \phi)]\right) = \frac{1}{4}
\end{equation}
And for the entangled history state $\frac{1}{\sqrt{2}} ([z^+] \odot [z^+] + [z^-] \odot [z^-])$ we likewise have
\begin{equation}
M\left(\frac{1}{\sqrt{2}} ([z^+] \odot [z^+] + [z^-] \odot [z^-])\right) = \frac{1}{4}
\end{equation}
Therefore, we see that the functional $M$ is not sufficient to distinguish between different quantum temporal structures.

We can improve the situation by defining a more discriminating functional, which is essentially the variance of $\left|\text{Proj}_{|\chi(\theta_1, \phi_1)\rangle, |\chi(\theta_2, \phi_2)\rangle} (\Psi) \right|^2$.  Specifically, we define
\begin{equation}
V(\Psi) := \frac{1}{(2 \pi)^4}\int_{[0, 2\pi]^4} \left(\left|\text{Proj}_{|\chi(\theta_1, \phi_1)\rangle, |\chi(\theta_2, \phi_2)\rangle} (\Psi) \right|^2 - \frac{1}{4} \right)^2 \, d\theta_1 \, d\phi_1 \, d\theta_2 \, d\phi_2
\end{equation}
Then we have
\begin{align}\label{VResult1}
V(\text{initial state }|\Psi(\theta, \phi)\rangle) &= \frac{115 + 25 \cos(4\theta)}{2048} \\
\label{VResult2}
V\left(\frac{1}{|\langle \Psi(\theta', \phi')|\Psi(\theta, \phi)\rangle|}\,[\Psi(\theta', \phi')] \odot [\Psi(\theta, \phi)]\right) &= \frac{57 + 11 (\cos (4\theta) + \cos (4 \theta')) + \cos (4\theta) \, \cos (4\theta')}{1024} \\
V\left(\frac{1}{\sqrt{2}} ([z^+] \odot [z^+] + [z^-] \odot [z^-])\right) &= \frac{3}{128}
\end{align}
Optimizing Eqn.\,(\ref{VResult1}), we find that
\begin{align}
\label{strongIneq1}
\frac{45}{1024} &\leq V(\text{initial state }|\Psi(\theta, \phi)\rangle) \leq \frac{35}{512}
\end{align}
where $\theta_{\text{min}} =  \pi/4$ and $\theta_{\text{max}} = 0$, and optimizing Eqn.\,(\ref{VResult2}) we obtain
\begin{align}
\label{strongIneq2}
\frac{9}{256} &\leq V\left(\frac{1}{|\langle \Psi(\theta', \phi')|\Psi(\theta, \phi)\rangle|}\,[\Psi(\theta', \phi')] \odot [\Psi(\theta, \phi)]\right) \leq  \frac{5}{64}
\end{align}
where $(\theta_{\text{min}},\theta_{\text{min}}') = (\pi/4, \pi/4)$ and $(\theta_{\text{max}},\theta_{\text{max}}') =  (0, 0)$.  Any state that falls outside both of the above inequalities must be an entangled history state.  For $\frac{1}{\sqrt{2}} ([z^+] \odot [z^+] + [z^-] \odot [z^-])$ we find $V\left(\frac{1}{\sqrt{2}} ([z^+] \odot [z^+] + [z^-] \odot [z^-])\right) = \frac{3}{128} < \frac{9}{256}$.   

Note that any history state which violates the inequality in Eq. (\ref{strongIneq2}) automatically violates the inequality in Eq. (\ref{strongIneq1}).   Therefore, in order to demonstrate that a history state $\Phi$ exhibits temporal entanglement, it is \textit{sufficient} to  demonstrate either that
\begin{equation}
V(\Phi) < \frac{9}{256} \quad \text{or} \quad V(\Phi) > \frac{5}{64}
\end{equation}
$V(\Phi)$ can be determined experimentally using the methods outlined in the preceding section.

\section{Conclusion}

We have presented an experimental framework to create and to measure entangled history states by post-selection and controlled operations exploiting auxiliary qubits coupled to the system of interest.  This method allows us to superpose radically different versions of ``what happened".  We have also explored how to test the $\tau$CHSH inequality on history states, and developed a set of inequalities that are sharp enough to distinguish history states exhibiting temporal entanglement.  Our thought-experiments seem to be within the capabilities of contemporary experimental physics.

\end{document}